\documentclass[preprint,aps,showpacs,nofootinbib]{revtex4}

\usepackage{graphicx}
\usepackage{epsfig,amssymb,amsmath,color}
\usepackage{epstopdf}

\usepackage[colorlinks=true,citecolor=blue,linkcolor=blue]{hyperref}

\newcommand{\beq}{\begin{equation}}
\newcommand{\eeq}{\end{equation}}
\newcommand{\bea}{\begin{eqnarray}}
\newcommand{\eea}{\end{eqnarray}}
\newcommand{\bear}{\begin{array}}
\newcommand {\eear}{\end{array}}
\newcommand{\bef}{\begin{figure}}
\newcommand {\eef}{\end{figure}}
\newcommand{\bec}{\begin{center}}
\newcommand {\eec}{\end{center}}

\newcommand{\la}{\left\langle}
\newcommand{\ra}{\right\rangle}

\begin{document}
\title{Natural and Multi-Natural Inflation in Axion Landscape}

\author{Tetsutaro Higaki}
\email{thigaki@post.kek.jp}
\affiliation{Theory Center, KEK, 1-1 Oho, Tsukuba, Ibaraki 305-0801, Japan}

\author{Fuminobu Takahashi}
\email{fumi@tuhep.phys.tohoku.ac.jp}
\affiliation{Department of Physics, Tohoku University, Sendai 980-8578, Japan}
\affiliation{Kavli Institute for the Physics and Mathematics of the
  Universe (WPI), Todai Institutes for Advanced Study, University of Tokyo,
  Kashiwa 277-8583, Japan}

\begin{abstract}
We propose a landscape of many axions,  where the axion potential receives various contributions
from shift symmetry breaking effects. We show that the existence of the axion with a super-Planckian decay constant
is very common in the axion landscape for a wide range of  numbers of axions and shift symmetry breaking terms,
because of the accidental alignment of axions. The effective inflation model is either natural or multi-natural inflation
in the axion landscape, depending on the number of axions and the shift symmetry breaking terms.
The tension between BICEP2 and Planck could be due to small modulations 
to the inflaton potential or steepening of the potential along the heavy axions after the tunneling. 
The total duration of the slow-roll inflation our universe experienced is not significantly larger than $60$ if the typical
height of the axion potentials is of order $(10^{16-17}{\rm \,GeV})^4$.
\end{abstract}
\preprint{KEK-TH-1730, IPMU14-0107, TU-967}
\maketitle

\section{Introduction}
The discovery of the primordial $B$-mode polarization of cosmic microwave background (CMB) 
by BICEP2 \cite{Ade:2014xna}  provides us with valuable information on the early universe. The measured tensor-to-scalar ratio
reads $r = 0.20^{+0.07}_{-0.05}$, which, 
taken at face value, implies large-field inflation models such as quadratic
chaotic inflation \cite{Linde:1983gd}, natural inflation \cite{Freese:1990rb}, or their extensions 
to polynomial \cite{Nakayama:2013jka,Nakayama:2013txa} or multi-natural inflation \cite{Czerny:2014wza}. 
In particular, the inflaton field excursion exceeds the Planck scale, which places a tight constraint
on the inflation model building. 

The observed large tensor-to-scalar ratio, however,
has a tension with the Planck results, $r < 0.11~(95\%{\rm ~CL})$ \cite{Ade:2013uln}. 
The tension can be relaxed if the scalar perturbations are suppressed at large scales~\cite{Miranda:2014wga},
which might be a result of the steep potential after the false vacuum decay via bubble 
nucleation~\cite{Freivogel:2014hca,Bousso:2014jca}. Alternatively, the tension can be relaxed 
by a negative running of the spectral index \cite{Ade:2014xna,Cheng:2014bta}. The running can be generated if there are 
small modulations to the inflaton potential~\cite{Kobayashi:2010pz,Czerny:2014wua}.
We shall see that the BICEP2 result and its apparent tension with Planck can be naturally
explained in a landscape of many axions.

In the landscape paradigm \cite{Bousso:2000xa, Susskind:2003kw}, there are numerous false vacua where
eternal inflation occurs, continuously producing universes by bubble nucleation. Our universe is considered to be 
within a single
bubble inside which slow-roll inflation took place after the tunneling event. This paradigm has various
implications for cosmology and particle physics and seems to have gained further momentum after BICEP2.
It however remains unanswered why and how the slow-roll inflation took place after the false vacuum decay. It might be
due to some fine-tuning of the potential. Such fine-tuning may be justified because,
most probably, there is an anthropic bound on the duration of slow-roll inflation after the bubble 
nucleation \cite{Freivogel:2005vv}. Still, it is uncertain how a very flat inflaton potential extending beyond the
Planck scale is realized in the landscape. Also, there is no clear connection
between the slow-roll inflation and the landscape paradigm.  
 
\begin{figure}[t!]
\begin{center}
\includegraphics[width=12cm]{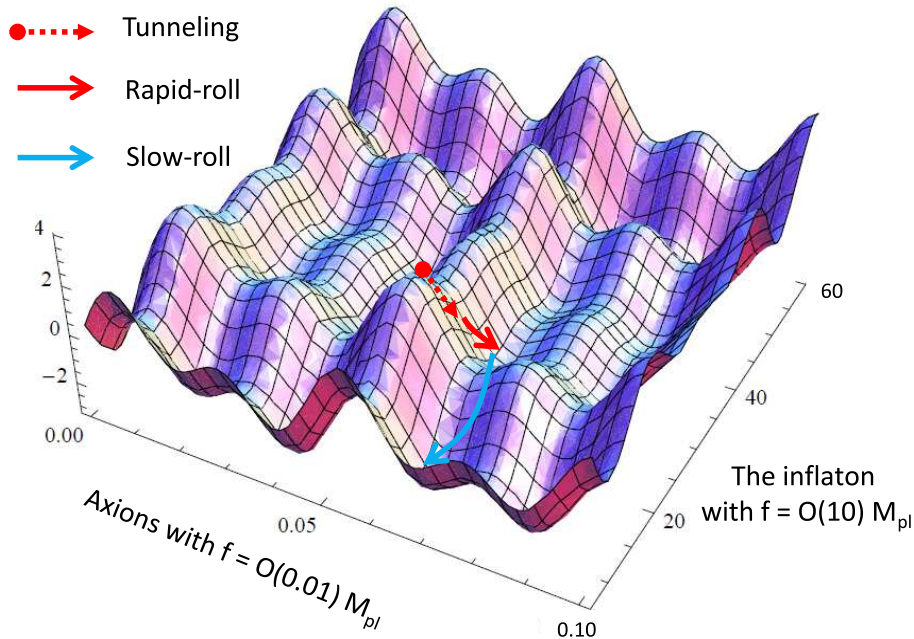}
\caption{
Illustration of the axion landscape.
The landscape consists of  many axions with  sinusoidal potentials of various height and periodicity.
There is likely to be a flat direction with an effective super-Planckian decay constant because of
the accidental alignment of axions, whereas the typical curvature at the false vacua is much larger
than the Hubble parameter.  The inflaton is one of the lightest axions, and the natural or multi-natural
inflation takes place after the last Coleman-De-Luccia tunneling event.
}
\label{FigLand}
\end{center}
\end{figure}

In this Letter we propose a landscape of  axions where the eternal inflation occurs in the false vacua and
the slow-roll inflation regime naturally appears after the tunneling events. See Fig.\ref{FigLand} for illustration of this concept. 
Most important, we find that there is very likely to be a direction along which the effective decay constant is super-Planckian, 
because of accidental alignment of axions known as the Kim-Nilles-Peloso (KNP) mechanism~\cite{Kim:2004rp}.\footnote{See also \cite{Harigaya:2014eta}.}
The important parameters that characterize the size and shape of the 
axion landscape are the number of axions, $N_{\rm axion}$, and that of shift symmetry breaking terms, $N_{\rm source}$.
In the case of $N_{\rm source} > N_{\rm axion}$,
the vacuum structure of the axions generates the so-called landscape, in which there are valleys and hills in the axion
potential; most important, there exist  many local vacua where the eternal inflation occurs, leading to a multiverse.
On the other hand, in the case  of $N_{\rm source} \leq N_{\rm axion}$, all the vacua are degenerate in energy,
and we need to embed the axion landscape into the string landscape to induce eternal inflation.  
In both cases the KNP mechanism works and the super-Planckian decay constant can be generated
from sub-Planckian ones. 
We will estimate the probability for obtaining a super-Planckian decay constant
for a wide range of values of $N_{\rm axion}$ and $N_{\rm source}$ based on a simplified model.
When $N_{\rm axion}=N_{\rm source}$, we will show that the enhancement of the
decay constant by a factor of $10^3$ happens with a probability of about $0.8\%$, 
$8\%$, and $24\%$ for $N_{\rm axion} = 10,100$ and $300$, respectively.  Thus, the existence of 
such a flat inflationary potential over super-Planckian field values is  a built-in feature of the axion landscape. 
We will also study the other cases;
when $N_{\rm source} > N_{\rm axion}$, 
it becomes less likely to obtain a large enhancement of the decay constant, whereas numerous local minima are generated.
When $N_{\rm source} < N_{\rm axion}$, there appear massless (or extremely light) axions, which may play
an important cosmological role.

Here let us mention the work relevant to the present study. Recently, it was shown in \cite{Choi:2014rja} that 
the KNP mechanism works for the case of $N_{\rm axion} = N_{\rm source} > 2$ with  smaller values of anomaly coefficients.
They estimated the probability for a given direction to have an effective super-Planckian decay constant,
and they found that it was roughly given by the inverse of the enhancement factor. Our result looks significantly larger than
their estimate, but this is because we have estimated a probability that the decay constant corresponding to {\it the lightest} direction 
happens to be super-Planckian for each configuration of the axion landscape. We obtained a consistent result 
 when we used the same values of  $N_{\rm axion} = N_{\rm source}$ adopted in \cite{Choi:2014rja}.

The inflation dynamics along the plateau is given by  either natural inflation \cite{Freese:1990rb} or multi-natural inflation \cite{Czerny:2014wza}, depending on $N_{\rm axion}$ and $N_{\rm source}$ as well as the height of each terms.
In a sufficiently complicated landscape with $N_{\rm source} \gtrsim N_{\rm axion}$, 
we expect that the latter will be more generic. As we shall discuss shortly,  the decay constant, inflaton mass, and duration of inflation etc. are determined by $N_{\rm axion}$ and $N_{\rm source}$,  i.e., the size and shape of the axion landscape.
The axion landscape thus provides a unified picture of the slow-roll inflation in the landscape paradigm.
In string theory, axions tend to be lighter than geometric moduli owing to gauge symmetries, and so,
the axion landscape can be thought of as a low-energy branch of the string landscape paradigm.
In this case, the role of the axion landscape is to generate an axion with a super-Planckian decay constant.


\section{Natural and Multi-Natural Inflation}
We must have a good control of the inflaton potential over more than about $10$ times the reduced Planck scale 
for successful  large-field inflation suggested by the BICEP2 result \cite{Ade:2014xna}. One way for realizing 
a sufficiently flat potential over  large field ranges is to impose a shift symmetry on the inflaton:
\begin{align}
\phi \to \phi + C,
\end{align}
where $\phi$ is an inflaton field and $C$ is a real transformation parameter. In the string theory
there appear many moduli fields through compactifications, and their axionic partners respect
such shift symmetry and therefore axions tend to be lighter than the moduli fields.
Thus, string axions are a good candidate for the inflaton.\footnote{ 
See, e.g., Refs.\cite{Grimm:2007hs,Kallosh:2007cc,Blumenhagen:2014gta,Grimm:2014vva} 
for attempts to realize natural inflation in the string theory.}

In order to have a graceful exit of the inflation, we need to break the shift 
symmetry. One plausible way is to break the continuous shift symmetry down to a discrete one
by adding a sinusoidal function:
\begin{align}
V(\phi) = \Lambda^4 \left(1-\cos\left(\frac{\phi}{f}\right)\right),
\label{natural}
\end{align}
where $f$ is the decay constant and $\Lambda$ denotes a dynamically generated scale violating the 
continuous shift symmetry.
Then, the natural inflation \cite{Freese:1990rb} takes place  for
a sufficiently large $f$. In particular,   $f \gtrsim 5 M_{Pl}$ is required by the Planck result \cite{Ade:2013uln}, 
and a similar bound is obtained by a combined analysis of BICEP2 and Planck \cite{Freese:2014nla},
where  $M_{Pl} \simeq 2.4 \times 10^{18}$\,GeV is the reduced Planck mass.

One can  extend the natural inflation to include multiple sinusoidal functions of different
height and decay constants;
\begin{align}
V(\phi) = \sum_i \Lambda_i^4  \cos\left(\frac{\phi}{f_i} + \theta_i \right) + V_0,
\label{multi-natural}
\end{align}
where $\theta_i$ denotes a relative phase between different sinusoidal functions
and a constant term $V_0$ is to make the cosmological constant vanish in the present vacuum.
This is the multi-natural inflation \cite{Czerny:2014wza,Czerny:2014xja,Czerny:2014qqa}.\footnote{
A similar potential with an irrational ratio of the decay constants was 
considered in \cite{Banks:1991mb}, and also recently in \cite{Kallosh:2014vja}.
It was also studied in the context of curavtons in \cite{Takahashi:2013tj}.
}
This potential is naturally generated if the axion is coupled to multiple gauge theories and/or stringy instantons 
\cite{Witten:1996bn,Blumenhagen:2009qh}. 
Note that the potential contains many local 
vacua \cite{Czerny:2014wza}.
The multi-natural inflation can accommodate a wide range of values of the spectral index $n_s$
and the tensor-to-scalar ratio $r$ \cite{Czerny:2014wza,Czerny:2014xja,Czerny:2014qqa}.
 If there is a mild hierarchy of order $10-100$ among the decay constants,
it is  possible to generate a sizable running of the spectral
index, $dn_s/d \ln k \simeq -0.03 \sim -0.02$, which remains approximately constant 
over the CMB scales without any conflict with large-scale 
structure data~\cite{Kobayashi:2010pz,Czerny:2014wua}. In the multi-natural inflation, 
at least one of the decay constants must be greater than the Planck scale to explain the BICEP2 result,
although the lower bound on $f_i$ is relaxed compared to the natural inflation.

\section{Kim-Nilles-Peloso mechanism}
The central issue for successful natural or multi-natural inflation 
is how to realize the effective decay constant greater than the Planck scale.
There is an interesting possibility to generate such a large decay constant from sub-Planckian 
decay constants, proposed by Kim, Nilles and Peloso \cite{Kim:2004rp}. Their idea is very simple.
Let us consider two axions with 
a shift symmetry, $\phi_i \to \phi_i + 2\pi f_i$.
We assume that the shift symmetry is broken by the following potential:
\begin{align}
V(\phi_1, \phi_2) = \Lambda_1^4  \cos\left(a_{11}\frac{\phi_1}{f_1} +a_{12} \frac{\phi_2}{f_2} \right) 
                           + \Lambda_2^4  \cos\left(a_{21}\frac{\phi_1}{f_1} +a_{22} \frac{\phi_2}{f_2}\right)  
 			  + V_0
\end{align}
where  $i,j$ run over $1,2$ and $a_{ij}$ is an integer-valued anomaly coefficient which depends on
the sources for the shift symmetry breaking. The typical value of the decay constants, $f_1$ and $f_2$,
are  considered to be around the string  scale $M_{\rm string} = {\cal O}(10^{17})$GeV.
The point is that, if $a_{11}/a_{12} = a_{21}/a_{22}$,
one combination of $\phi_1$ and $\phi_2$ does not appear in the scalar potential, and therefore, 
there is a flat direction.
It implies that, if $a_{11}/a_{12} \approx a_{21}/a_{22}$
but $a_{11}/a_{12} \ne a_{21}/a_{22}$, the flat direction is lifted, and its corresponding decay constant 
can be much larger than the typical value of $f_i$.
 In other words, a certain alignment of the axions appearing in the
sinusoidal functions can  generate a super-Planckian decay constant from sub-Planckian ones.
The required enhancement is of order $10^{2-3}$ for the decay constant around
the string scale. 

The KNP mechanism is very useful to embed the natural and multi-natural inflation in UV theory
such as string theory where the typical values of the decay constant is below the Planck scale. For instance,
multi-natural inflation in string-inspired setting was discussed in \cite{Czerny:2014xja,Czerny:2014qqa} based on the 
KNP mechanism. Recently, the extension of the KNP mechanism to multiple axions was proposed in \cite{Choi:2014rja},
where they showed that the KNP mechanism works in the presence of multiple axions, and they estimated
the probability to realize an effective large decay constant, $f_{\rm eff}$ in the case of $N_{\rm axion} = N_{\rm source}$,
assuming that the other directions are sufficiently heavy.
They showed that the  probability for the alignment of axions is approximately given by the inverse 
of the enhancement $f_{\rm eff}/f_i$. 
Later in this letter we will directly calculate the distributions of the axion mass 
for a  wide range of values of $N_{\rm axion}$ and  $N_{\rm source}$, based on a simplified model of the axion landscape.
There we will show that the required enhancement can be generically realized in the axion landscape.

\section{False vacuum decay}
In the landscape paradigm, eternal inflation occurs in numerous false vacua, creating 
various universes through the tunneling, leading to a multiverse.  As pointed out in Ref.~\cite{Freivogel:2005vv},
if we live in a bubble created from the false vacuum decay followed by the slow-roll inflation with 
the e-folding number $N_e = 50 - 60$, we may be able to observe
a negative curvature as a remnant of the bubble nucleation. The detection of the negative spatial curvature
will be more likely if there is a pressure toward shorter inflation.  Also, if the inflaton potential after the bubble nucleation is sufficiently steep,  
 scalar density perturbations at large scales can be suppressed~\cite{Yamauchi:2011qq,Bousso:2013uia}.  
 This could explain the low $\ell$-anomaly of CMB. 
 After BICEP2,  the problem of the low-$\ell$ suppression was sharpened~\cite{Bousso:2014jca}.
Recently, the landscape and its implications were  also discussed in the sneutrino chaotic inflation model~\cite{Murayama:2014saa}.
 As we shall see shortly,  these expected properties are built-in features of the axion landscape.

The Coleman-De-Luccia (CDL) instanton~\cite{Coleman:1980aw} requires a rather large curvature of the potential,
$V'' \gg H^2$ in the false vacuum; otherwise the tunneling occurs as in the Hawking-Moss case \cite{Hawking:1982my}.
In order to realize both the eternal inflation in the false vacuum, CDL instanton, and the subsequent 
slow-roll inflation in terms of a single scalar field, one needs a rather
contrived functional form of the inflaton potential. If one considers multiple fields, on the other hand, we can naturally 
realize mass hierarchy between the false vacuum and the slow-roll inflation regime \cite{Linde:1995xm}. 
Still, the existence of the inflationary plateau had no direct relation with the landscape so far.
This questions is also answered naturally in the axion landscape.

\section{Axion Landscape}
Let us now consider an axion landscape. For simplicity 
 we will focus on a landscape of  axions with sub-Planckian decay constants,
but the application to other cases is straightforward. 
The shift symmetries of axions are explicitly broken by various sources such as gaugino condensation and/or instanton:
\begin{align}
V(\phi_i) = \sum_{i=1}^{N_{\rm source}} \Lambda_i^4 \cos\left(\sum_{j=1}^{N_{\rm axion}} a_{ij} \frac{\phi_j}{f_j} + \theta_i\right) + V_0
\label{al}
\end{align}
with $\phi_i$ is an axion, $a_{ij}$ integer-valued anomaly coefficients,
 and $f_i$ the decay constant.\footnote{It is possible to add another functional form of the 
 axion potential \cite{Silverstein:2008sg,McAllister:2008hb,Kaloper:2008fb}, but it does not
 significantly change our arguments, as long as there are sufficient number of axions which form the landscape like (\ref{al}).
In fact, such an additional potential will help to generate many false vacua.
  }
 The axion is assumed to have a periodicity,
 \begin{align}
 \phi_i \to \phi_i + 2 \pi f_i.
 \end{align}
The potential height $\Lambda_i$ is naively expected to be of order 
$M_{\rm string} = {\cal O}(10^{17})\,{\rm GeV}$ in the string theory, whereas 
some of them can be smaller if they are dynamically generated. 
In general, $N_{\rm source}$ does not necessarily coincide with $N_{\rm axion}$.
For the moment we assume $N_{\rm source}$ is comparable to  $N_{\rm axion}$,
and we shall discuss the other cases and its cosmological implications later.
Here and in what follows, $N_{\rm source}$  counts only the shift symmetry breaking terms which can affect the vacuum structure
and/or the inflaton dynamics, and hierarchically small contributions are to be considered separately in the low energy.
For a sufficiently large $N_{\rm source}$, 
they form a complicated landscape with numerous local minima,
where eternal inflation occurs, continuously creating new universes by bubble nucleation (cf. Fig.~\ref{FigLand}).
On the other hand, when $N_{\rm source}$ is equal to or smaller than $N_{\rm axion}$,
all the vacua are degenerate in energy. This is not an obstacle for generating a direction with a super-Planckian
decay constant. In order to induce eternal inflation, however, we would need to either introduce another 
kind of shift symmetry breaking to lift the degeneracy or
embed the axion landscape into the string landscape which contains many local vacua.
In the latter case, the role of the axion landscape is to provide an axion with the super-Planckian decay constant.


In the axion landscape, 
a super-Planckian decay constant for the inflaton is likely realized by accidental alignment
of axions in the axion landscape. As will be shown below, the probability to obtain an enhancement of
the decay constant by a factor of $10^3$ is about $1\% (10\%)$ for $N_{\rm axion} = N_{\rm source} = 10 (100)$.
On the other hand, the probability decreases as $N_{\rm source}$
becomes larger than $N_{\rm axion}$. The case of $N_{\rm axion} = N_{\rm source}$ was recently 
studied in detail in \cite{Choi:2014rja}, and they showed that  the KNP mechanism works for multiple axions
with smaller values of the anomaly coefficients. 
Contrary to  \cite{Choi:2014rja}, however, we believe that no hierarchy in the potential height is needed to
ensure the mass hierarchy between the inflaton and the other axions. This has an important implication
for the duration of the inflation as we discuss later. 
Thus the KNP mechanism generically works for a wide range of values of  $N_{\rm axion}$, $N_{\rm source}$,
and the potential height $\Lambda_i^4$.  The slow-roll inflation is a natural outcome of the axion landscape.

If the axion landscape has many local vacua, the CDL instantons are formed at the tunneling.
This is because the typical decay constants are sub-Planckian and
the axions other than the inflaton tend to be so heavy that the barrier
between the local minima is narrow.
After the tunneling, the universe will be dominated by the curvature term and the heavy axions for a while. 
The slow-roll inflation by the lightest axion starts when its energy density dominates the universe. 

In order to get the feeling of the probability for a sufficiently flat direction to arise by the accidental alignment, let us consider
a simplified model. We set $\Lambda_i = \Lambda$, $f_i =f$ and $\theta =0$ in Eq.~(\ref{al}).\footnote{
In general, some of the phases are physical and cannot be absorbed by the redefinition of the parameters or
the shift of axions. In fact, they will play an important role in the multi-natural 
inflation~\cite{Czerny:2014wza,Czerny:2014xja,Czerny:2014qqa} to obtain a wide range of values of $n_s$ and $r$.
} We take $a_{ij}$ as an integer valued random matrix satisfying $-n \leq a_{ij} \leq n$ to scan various realization of the
axion landscape. In order to find the flattest direction, we Taylor expand the cosine function up to the second order of
axions. Then  the mass matrix for axions $\{\phi_i\}$ is proportional to
\begin{align}
M_{ij}^2 \;\propto\; A_{ij} \equiv \sum_{k=1}^{N_{\rm source}} a_{ki} a_{kj}.
\end{align}
Of course, as the field values change, the contribution of each shift symmetry breaking term to the mass term changes.
Since we are interested in the case that all the axions other than the lightest one are stabilized in one of the false vacua,
such an expansion can be justified. 

We numerically estimate the eigenvalue distribution of the integer-valued random matrix squared, $A_{ij}$.
Let us denote the eigenvalues of $A_{ij}$ as $0< a_1^2 \leq a_2^2 \leq \cdots \leq a^2_{N_{\rm axion}}$.\footnote{
We focus on a situation that $a_1$ is smaller than $a_2$, and identify the lightest axion with the inflaton. 
On the other hand, assisted inflation \cite{Liddle:1998jc} or N-flation \cite{Dimopoulos:2005ac} will be possible
if some of the eigenvalues are degenerate. 
} 
Here we exclude the case of ${\rm det} A = 0$, in which case there is always massless axion(s).  This should be included
in the case of $N_{\rm source} < N_{\rm axion}$, as some of the source terms are not independent (as far as the axion mass
is concerned). The effective decay constant for each axion is enhanced by $\la a \ra/a_i$ with respect to its typical value $f$.
For instance, the effective decay constant for the lightest axion is given by
 \begin{align}
 f_{\rm eff} \simeq f_i \frac{\la a \ra}{a_1},
 \label{eq:rel}
 \end{align}
 and similarly for the second or third lightest axions.
Here $\la a \ra \equiv  \sqrt{\la a_i^2  \ra}$ is the squared root of the averaged eigenvalues of $A_{ij}$. 
Note that $\la a \ra$ scales as $\sqrt{N_{\rm axion}}$, which however does not change the relation (\ref{eq:rel}) as it only
affects the typical mass scale, not the decay constants.

Before going to the detail, we give the result on the integrated probability  that
the  effective decay constant for the lightest axion is  enhanced by more than $f_{\rm eff}/f_i$, 
for the case of $N_{\rm source} = N_{\rm axion}$:
\begin{align}
{\cal P} (f_{\rm eff}/f_i)& \sim N_{\rm axion}\left(\frac{f_i}{f_{\rm eff}}\right)
\end{align}
%
%
as long as it is much smaller than unity. Here ${\cal P} (x)$ denotes the probability that the enhancement
factor $f_{\rm eff}/f_i$ for the lightest axion becomes larger than $x$.
 This implies that the the effective decay constant for the lightest axion
is enhanced as $N_{\rm axion}$ increases.

\begin{figure}[t!]
\begin{center}
\includegraphics[width=12cm]{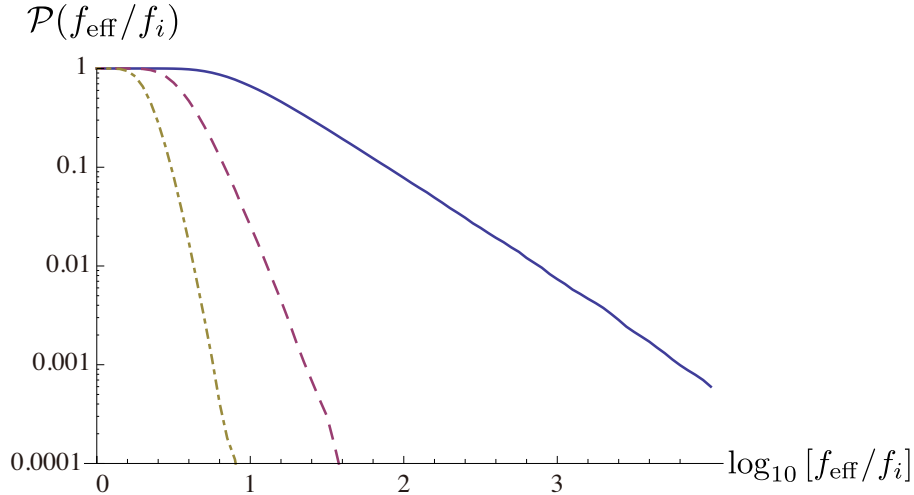}
\caption{
The integrated probability distribution functions, ${\cal P}(f_{\rm eff}/f_i)$, for the
lightest (solid), second lightest (dashed), and third lightest (dash dotted) mass eigenvalues, from right to left.
We generated $10^5$ random matrices with $N_{\rm axion} = 10$ and $n=2$. The probability for
the enhancement by more than $10^3$ is about $1\%$. 
It is rare that two (or three) flat directions arise simultaneously by the accidental alignment.
}
\label{fig:IPD1}
\end{center}
\end{figure}

 In Fig.~\ref{fig:IPD1} we show the integrated probability distribution functions for 
 the enhancement factor of the effective decay constant for the lightest, the second lightest, and the third lightest
 axions. We have generated  $10^5$ random matrices $A_{ij}$ with ${\rm det} A \ne 0$ for $N_{\rm axion}=10$ and $n=2$.
For instance we can see that the probability that the lightest axion
mass happens to be $10^3 (10^2)$  times lighter than the averaged mass is about $0.8 \% ( 8\%)$.
In other words, the decay constant for the lightest axion is enhanced by a factor of $10^3$ than the typical value
with a probability of about $1\%$ for $N_{\rm axion}=10$. Surprisingly, even for a relatively small axion landscape that consists of
only $10$ axions, we can realize the enhancement of $10^3$ with a probability about $1\%$.\footnote{
We are aware that it is non-trivial to interpret  the probability because of the measure problem.}
 As one can see from the figure, it is much more unlikely to have two (or more) relatively flat directions simultaneously. 
This suggests that our inflaton is the lightest axion in the axion landscape. 

Note that our result on the probability
is about one order of magnitude larger than the result of Ref.~\cite{Choi:2014rja}. This is because they estimated a probability
for a given combination of axions to have a super-Planckian decay constant 
assuming the rest of axions are hierarchically heavier. Instead, we have evaluated
all the mass eigenvalues for each realization of the axion landscape, and then estimated the probability for {\it the lightest} one to have
super-Planckian decay constants. This corresponds to the probability that, for a given axion landscape, we find at least one flat
direction with the super-Planckian decay constant. 
 In addition, we can easily extend our analysis to the case of $N_{\rm source} > N_{\rm axion}$
as we directly estimate the mass eigenvalue distribution.

\begin{figure}[t!]
\begin{center}
\includegraphics[width=12cm]{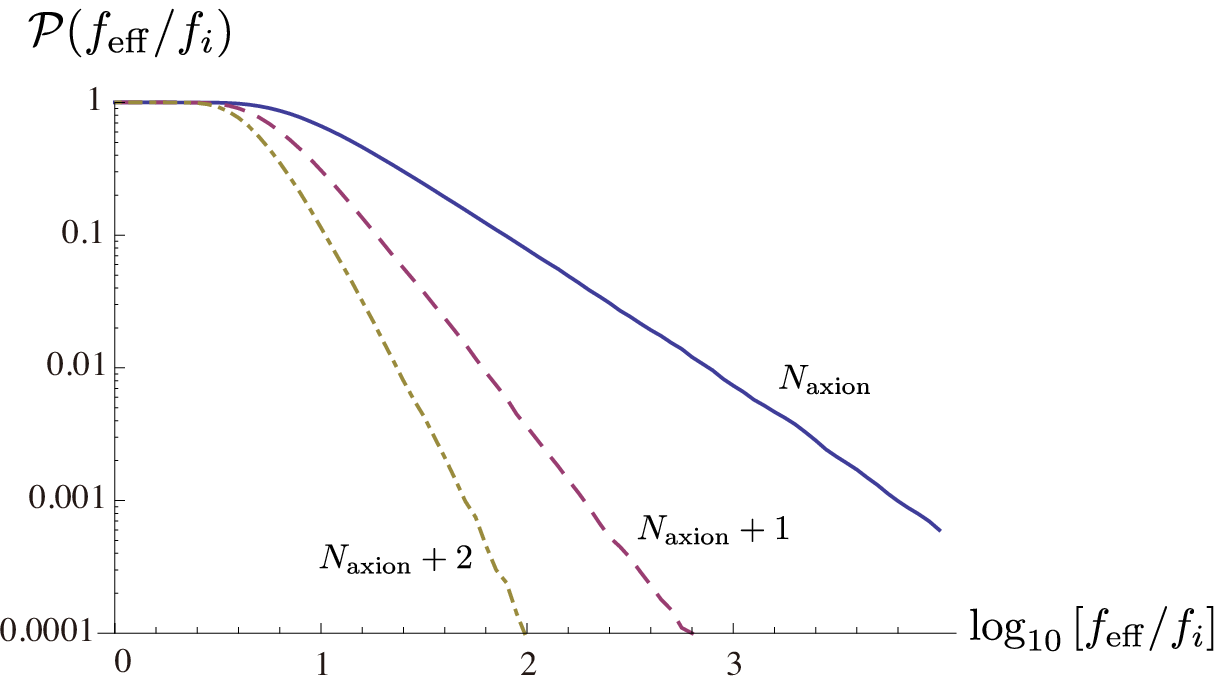}
\caption{
The integrated probability distribution functions, ${\cal P}(f_{\rm eff}/f_i)$, for the case of
$N_{\rm source} = N_{\rm axion}, N_{\rm axion}+1, N_{\rm axion}+2$, from right to left.
We generated $10^5$ random matrices and set $N_{\rm axion} = 10$.
As the number of shift symmetry breaking terms increases, it becomes more difficult
to realize the suppression of the lightest mass eigenvalues (or equivalently, the enhancement of the decay constant).
}
\label{fig:IPD1b}
\end{center}
\end{figure}

We show in Fig.~\ref{fig:IPD1b} the integrated probability distribution function for
$f_{\rm eff}/f$ in the case of $N_{\rm source} =  N_{\rm axion}, N_{\rm axion}+1, N_{\rm axion}+2$. As the number of source
terms increases, it becomes more difficult to have a light axion with a super-Planckian decay constant. 
Roughly, the probability distribution scales as 
${\cal P}_{N_{\rm axion}+1} \sim ({\cal P}_{N_{\rm axion}})^2$ and  ${\cal P}_{N_{\rm axion}+2} \sim ({\cal P}_{N_{\rm axion}})^3$.
This can be intuitively understood as follows. Note that,  if one replaces one of the rows of the random matrix $A_{ij}$ 
with another random-valued  row, the resultant matrix will be almost statistically independent from the initial one.
Therefore, adding another source term is approximately equivalent to considering another sets of axions with 
$N_{\rm axion} = N_{\rm source}$. In order to have a suppression of the axion mass (or equivalently, the enhancement of the decay constant),
the accidental cancellation must happen in the two sets of axions simultaneously. Therefore, we expect 
${\cal P}_{N_{\rm axion}+1} \sim ({\cal P}_{N_{\rm axion}})^2$, and similarly for a larger value of the source terms.

\begin{figure}[t!]
\begin{center}
\includegraphics[width=10cm]{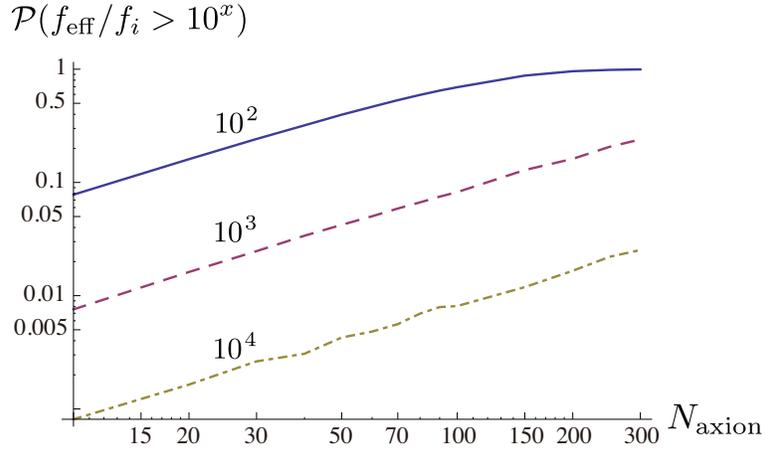}
\caption{
The probability to have a flat direction with the decay constant
more than $10^2$ (upper), $10^3$ (middle), and $10^4$ (bottom) times larger than the typical value in one realization of the
axion landscape, as a function of  $N_{\rm axion} =  N_{\rm source}$.
}
\label{fig:IPD2}
\end{center}
\end{figure}

Now let us go onto a larger axion landscape. To simplify our analysis we focus on the case of $N_{\rm axion} =
N_{\rm source}$. We show in Fig.~\ref{fig:IPD2} the probability to have a flat direction with the decay constant
more than $10^2$ (upper), $10^3$ (middle), $10^4$ (bottom)  times larger than the typical value in one realization of the axion landscape, 
as a function of  $N_{\rm axion}$.
We can see that the probabilities for the enhancement of $10^2$,  $10^3$, and $10^4$
are about $70\%$,  $8\%$, and $0.8\%$, respectively for $N_{\rm axion} \approx 100$, and they  further increase in proportion 
to $N_{\rm axion}$. Note that the enhancement by a factor of $10^4$ is less likely unless
 the number of axions becomes much larger than ${\cal O}(100)$. 
If we compare our result at $N_{\rm axion} = 10$ with that of \cite{Choi:2014rja}, ours is about one order of magnitude
larger, probably because we focus on the lightest direction; if we randomly choose one direction from the $10$ directions, 
the probability to realize the enhancement decreases by a factor of $10$. 
 Thus, we conclude that the existence of the axion with a super-Planckian decay constant
 is very common in the axion landscape. A flat potential over the super-Planckian field values emerges from
 a complex vacuum structure in the axion landscape. The likely size of the decay constant sensitively depends on
 $N_{\rm axion}$ and $N_{\rm source}$.

The effective potential for the inflaton generically contains  a combination of various
sinusoidal functions. 
Note that the height of the inflaton potential is generically comparable to  the typical height 
in the axion land scape, although it can be suppressed if there is a hierarchy in the size of the
shift symmetry breaking. This is the case if some of them are dynamically generated. 
If the inflaton potential is dominated by a single sinusoidal function, the inflaton dynamics
is well approximated by the natural inflation (\ref{natural}). On the other hand, if there are multiple sinusoidal
functions with slightly different decay constants and heights, it will be the multi-natural inflation (\ref{multi-natural}). 
This is the case if $N_{\rm source} > N_{\rm axion}$.
In particular,  if $N_{\rm source}$ is slightly larger than $N_{\rm axion}$ and some of 
the source terms have hierarchically small height,  the inflaton potential receives small modulations, 
leading to a sizable negative running of the spectral index \cite{Kobayashi:2010pz,Czerny:2014wua}.
Thus, the running of the spectral index, if detected, will constrain the relation between $N_{\rm source}$
and $N_{\rm axion}$.

Now let us consider the eternal inflation which occurs in the false vacua before the slow-roll inflation takes place. 
The tunneling rate is not sensitive to the value of the inflaton,
 since the couplings between the inflaton and the heavy axions are weak \cite{Linde:1995xm}.
Therefore, the inflaton field value at the tunneling is generically deviated from the inflaton potential minimum by ${\cal O}(f_{\rm eff})$,
which determines the typical duration of the slow-roll inflation.\footnote{
If the energy difference is small, the inflaton might be more or less stabilized near the potential minimum.
This situation can be avoided if the typical energy density in the false vacua is larger than the inflaton potential
height.
}
The mass hierarchy between the false vacuum and the inflation plateau
is also determined by $f_{\rm eff}/f_i$, the value of which depends on the detailed properties of the axion landscape. 
It is a quantitative question how large hierarchy is generated or how long inflation typically lasts, but,
based on our simple model of the axion landscape, 
it is likely  that the decay constant $f = {\cal O}(10) M_{Pl}$ is generated from the typical decay constants
$f_i = {\cal O}(10^{16-17})$\,GeV with $N_{\rm axion} \sim N_{\rm source}= {\cal O}(10-100)$, 
and the typical e-folding number after
the tunneling event is not significantly larger than  $60$.
Intriguingly, the BICEP2 result suggests the inflation scale to be
$V_{\rm inf} \simeq (2.0 \times 10^{16}\,{\rm GeV})^4\cdot (r/0.16)$, 
not far from $M_{\rm string}^4 = {\cal O}((10^{17}\,{\rm GeV})^4)$, 
the potential height at false vacua naively expected 
in the string theory.  The two scales can be even closer if the axion potentials are dynamically generated.
Since the two scales are comparable, it implies that the inflaton field value at the horizon exit of cosmological scales 
is of order $f_{\rm eff}$, which should be of order $10$ times the Planck scale, $M_{Pl}$.
That is to say, the typical duration of the inflation is not significantly different from $60$.

It sensitively depends on the duration of inflation whether we will be able to observe the negative spatial curvature.
The detection is more likely if there is a pressure toward shorter inflation. In the axion landscape, this is generically
the case because the flat direction arises from the accidental alignments.  In particular,  the pressure toward shorter inflation will be significant if  the number of axions is relatively small,
or if the number of source terms tends to be (much) larger than the number of axions. On the other hand,
the pressure might be weakened if there are many axions with $N_{\rm axion} \sim N_{\rm source}$, since
the large effective decay constant can be generated more easily. 
%
Therefore, detection or non-detection of the negative spatial curvature by future observations will provide us with
information on the size and shape of the axion landscape.


So far we have assumed $N_{\rm source} \sim N_{\rm axion}$. If $N_{\rm source} \gtrsim N_{\rm axion}$,
it will be more difficult to realize a super-Planckian decay constant, as we have seen in Fig.~\ref{fig:IPD1}.
For $N_{\rm source} < N_{\rm axion}$, 
there appear massless (or extremely light) $N_{\rm axion} - N_{\rm source}$ axions in the low energy,
while the above discussion on the inflation remains unchanged. 
They may behave as dark radiation or hot dark matter, ameliorating the tension between BICEP2 and 
Planck~\cite{Giusarma:2014zza,Zhang:2014dxk}.
They are produced through various mechanisms, e.g., coherent oscillations, thermal production, and non-thermal
production from decays of long-lived states \cite{Cicoli:2012aq,Higaki:2012ar,Higaki:2013lra}. 
If they get tiny masses, stable axion dark matter can generate isocurvature fluctuations,
which is strongly constrained by the BICEP2 result \cite{Higaki:2014ooa,Marsh:2014qoa,Visinelli:2014twa}. 
If there exist axions with masses at (more than) PeV scales,
they may play a role in generating fluctuations as curvatons \cite{Kawasaki:2012gg}.

Another interesting implication of the axion landscape is that the decompactification problem during the inflation
will be naturally evaded 
because the axions are the lightest moduli fields in the string theory owing  to the shift symmetry.
Although it is a non-trivial task to infer the supersymmetry breaking scale in the axion landscape,
there is a lower bound on the gravitino mass, $m_{3/2} \gtrsim H_{\rm inf} \sim 10^{14}\,{\rm GeV}$ \cite{Kallosh:2004yh},
in a known set-up of moduli stabilization. Here $m_{3/2}$ is the gravitino mass, which is typically comparable 
to other heavy moduli masses. Note that there exist moduli fields heavier 
than the gravitino, which do not play any cosmological role in our context.
In fact, such a high supersymmetry breaking is consistent with the Standard Model Higgs 
mass \cite{Hebecker:2012qp,Ibanez:2012zg,Hebecker:2013lha}.

\section{Reheating}

For successful inflation, the inflaton energy is transfered into the Standard Model particles,
and the baryon asymmetry is generated after the inflation.
To this end, we shall discuss the reheating temperature and the leptogenesis after the inflaton decay.
The mass of the inflaton $m_{\phi}$ in the natural or multi-natural inflation is given by
\begin{align}
m_{\phi} \sim \frac{\Lambda^2}{f_{\rm eff}} \sim 10^{13}~{\rm GeV},
\end{align}
where $\Lambda = {\cal O}(10^{16}){\rm\,GeV}$ and the effective decay constant $f_{\rm eff} = {\cal O}(10)M_{Pl}$.
Since the inflaton is one of the lightest axions, there exists a coupling  to the Stanard Model gauge bosons:
${\cal L} = c(\phi/f_a)F_{\mu \nu}\tilde{F}^{\mu \nu}$. The decay width of the inflaton reads
\begin{align}
\Gamma_{\phi} = N_g\frac{c^2}{4\pi}\frac{m_{\phi}^3}{f_a^2},
\end{align}
where $N_g = 8 + 3 + 1$ counts the number of Standard Model gauge bosons.
Note that the decay constant $f_a$ can be different from the those in the scalar potential. 
Then, the reheating temperature $T_R$ is estimated as
\begin{align}
T_R \sim 4 \times 10^{10}\, {\rm GeV}\cdot 
\left(\frac{m_{\phi}}{10^{13}~{\rm GeV}}\right)^{3/2} 
\left(\frac{f_a/c}{10^{17}~{\rm GeV}}\right)^{-1}.
\end{align}
Here, we used $T_R = (90/\pi^2 g_*(T_R)) \sqrt{\Gamma_{\phi}M_{\rm Pl}}$ and $g_*(T_R) = 106.75 $ counts 
the number of degree of the relativistic particles in plasma. 
Therefore successful thermal leptogenesis \cite{Fukugita:1986hr} is possible when 
the right-handed neutrinos is produced from the thermal bath:
\begin{align}
\frac{n_B}{s}\bigg|_{\rm max.} \sim 10^{-10}\cdot \left(
\frac{M_{N_1}}{10^{10}\,{\rm GeV}}
\right)
~~~{\rm for}~
M_{N_1} \lesssim T_R ,
\end{align}
where 
$M_{N_1}$ is the mass of the lightest right-handed neutrino.

For $T_R \lesssim M_{N_i} \lesssim m_{\phi}/2$,  non-thermal leptogenesis will be also possible,
if there exists a direct coupling to the right-handed neutrinos, say, ${\cal L} \sim (\phi/f_a) M_{N_i} N_i N_i$,
where $N_i$ is the right-handed neutrinos.
Then, the partial decay width reads
\begin{align}
\Gamma_{\phi}^{N} \simeq \frac{1}{16\pi} 
\left(\frac{M_{N_i}}{f_a}\right)^2
m_{\phi}
\end{align}
and the reheating temperature does not change significantly.
The right amount of the baryon asymmetry is then obtained through the decay of $N_i$ produced from the inflaton
and the sphaleron process:
\begin{align}
\frac{n_B}{s}\bigg|_{\rm max.} \sim 5 \times 10^{-10} 
\cdot 
\left( \frac{T_R/m_{\phi}}{10^{-3}}\right)
\left( \frac{M_{N_i}}{10^{12}\,{\rm GeV}}\right)
\left( \frac{Br}{10^{-2}}\right).
\end{align}
Here, 
$Br$ is the branching fraction of the decay mode of $\phi \to 2 N_i$.

\section{Conclusions}
We have proposed a landscape of many axions where the axion potential receives various contributions
from shift symmetry breaking effects. If the number of the shift symmetry breaking, $N_{\rm source}$, 
is large enough, there are valleys and hills in the axion potential; eternal inflation occurs in the local
minima,
continuously creating new universes via the CDL tunneling. On the other hand, if $N_{\rm source} \leq N_{\rm axion}$,
all the vacua are degenerate in energy. In this case, eternal inflation will be possible if one introduces another kind
of shift symmetry breaking or if one embed the axion landscape into the string landscape with a large number
of local minima. 

Interestingly, there is very likely to be a direction along which the effective decay constant exceeds the Planck scale owing to the accidental alignment of axions, i.e., the KNP mechanism. Therefore, in the axion landscape,
the existence of the slow-roll inflation regime is a natural outcome of the complicated vacuum structure. 
We have also  argued that the effective inflation model in the axion landscape will be either natural inflation 
or multi-natural inflation, depending the values of $N_{\rm axion}$ and $N_{\rm source}$.  In the latter case, a wide range of $(n_s, r)$
as well as the running of the spectral index can be realized. 
The size and shape of the axion 
landscape are parametrized by $N_{\rm axion}$ and $N_{\rm source}$, which determines the effective super-Planckian decay constant, and therefore the typical duration of the slow-roll inflation. In a certain case,
there might be a strong pressure toward shorter inflation, and it will be more likely that
we can measure the negative spatial curvature as a remnant of the CDL tunneling. 
Conversely,  non-detection of the negative curvature will constrain the size and shape of the axion landscape. 
Also, the size of density perturbations and the inflaton potential height will be useful to extract information on the axion
landscape.  A more quantitative study on this issue will be given elsewhere.

{\it Note added}: After completing this work, there appeared the papers \cite{Tye:2014tja, Kappl:2014lra, Ben-Dayan:2014zsa, Long:2014dta}, in which the alignment of two axions was discussed based on the KNP mechanism as well. We also note that the charge assignment studied in Refs.\cite{Tye:2014tja, Ben-Dayan:2014zsa}
leads to the suppression of the lighter axion mass, similarly to the seesaw mechanism for the light neutrino mass.

\section*{Acknowledgement}
This work was supported by 
Young Scientists (B) (No.24740135) [FT],   Scientific Research on Innovative Areas (No.23104008 [FT]),
and Scientific Research (B) (No.26287039 [FT]  and No. 25800169 [TH]),  by Inoue Foundation for Science [FT], and by World 
Premier International Center Initiative (WPI Program), MEXT, Japan [FT].

\bibliography{references}
\end{document}